\documentclass[11pt,a4paper,final]{article}

\usepackage{amsmath}
\usepackage{amsfonts}
\usepackage{amssymb}

\usepackage[retainorgcmds]{IEEEtrantools}

\usepackage{multirow}

\usepackage{graphicx}
\usepackage{tikz}

\usepackage{cite}

\def\a{{\alpha}}      %%%% Undotted Greek Indices %%%%
\def\b{{\beta}}
\def\g{{\gamma}}
\def\d{{\delta}}
\def\r{{\rho}}

\def\e{{\epsilon}}

\def\ad{{\dot{\alpha}}}  %%%% Dotted Greek Indices %%%%
\def\bd{{\dot{\beta}}}
\def\gd{{\dot{\gamma}}}

\def\rd{{\dot{\rho}}}

\def\D{{\rm D}}         %%%% Operators / Equations %%%%
\def\Dd{{\bar{\rm D}}}
\def\pa{\partial}

\def\[{\left[}
\def\]{\right]}

\def\be{\begin{equation}}
\def\ee{\end{equation}}
\def\bea{\begin{IEEEeqnarray*}}
\def\eea{\end{IEEEeqnarray*}}

\parskip=\medskipamount                 % space between paragraphs (LaTeX)
\thicklines                         % thick straight lines for pictures (LaTeX)

\font\ro=cmsy10                          % font with rope
\def\kcr{{\hbox{\ro \char'170}}}        % right-handed rope
\def\ktl{{\hbox{\ro \char'170}}}        % top end for left-handed rope
\def\ktr{{\hbox{\ro \char'170}}}        % " right
\def\kbl{{\hbox{\ro \char'170}}}        % " bottom left
\def\kbr{{\hbox{\ro \char'170}}}        % " right
\def\border{                           %%%% Border %%%%
        \setlength{\unitlength}{1mm}
        \newcount\xco
        \newcount\yco
        \xco=-21
        \yco=12
        \begin{picture}(140,0)
        \put(\xco,\yco){$\ktl$}
        \advance\yco by-1
        {\loop
        \put(\xco,\yco){$\kcr$}
        \advance\yco by-2
        \ifnum\yco>-250
        \repeat
        \put(\xco,\yco){$\kbl$}}
        \xco=158
        \yco=12
        \put(\xco,\yco){$\ktr$}
        \advance\yco by-1
        {\loop
        \put(\xco,\yco){$\kcr$}
        \advance\yco by-2
        \ifnum\yco>-250
        \repeat
        \put(\xco,\yco){$\kbr$}}
        \put(-20,13){\tiny **University of Maryland * Center for String and
         Particle  Theory* Physics Department***University of Maryland *Center
        for String and Particle  Theory** }
        \put(-20,-251.5){\tiny **University of Maryland * Center for String and
         Particle  Theory* Physics Department***University of Maryland *Center
        for String and Particle  Theory** }
        \end{picture}
        \par\vskip-8mm}

\def\endtitle{\end{quotation}\newpage}  %%%% End title page %%%%

\begin{document}

%%%%%%%%%%%%%%%%%%%%%%%%%%%%%%%%%%%%%%%%%%%%%%
%%%%  Title page %%%%%%%%%%%%%%%%%%%%%%%%%%%%%

%\border\headpic {\hbox to\hsize{\today % date    %%%% Boarder - Heading %%%%
%\hfill{Institution-xxx-xxx % Internal pre-print number
%}}}
%\par
%{$~$ \hfill
%{hep-th/xxxx.xxxx % hep-th preprint number
%}}
%\par

%%%%%%%%%%%%%%%%%%%%%%%%%%%%%%%%%%%%%%%%%%%%%%

\setlength{\oddsidemargin}{0.3in}     %%%% Title-Authors-Institution-Grant info %%%%
\setlength{\evensidemargin}{-0.3in}
\begin{center}
\vglue .10in
{\Large\bf {Linearized Non-Minimal Higher Curvature Supergravity}
% \footnote
%{Supported in part by %National Science Foundation Grant PHY-09-68854.
%}
}
\\[.5in]

\large{Fotis Farakos$^{a}$, Alex Kehagias$^{b,c}$ and Konstantinos Koutrolikos$^{b}$}
\\[0.2in]

%\vspace{.3cm}
{\normalsize {\it  $^{a}$ Institute for Theoretical Physics, Masaryk University, \\611 37 Brno, Czech Republic}}\\

\vspace{.3cm}
{\normalsize {\it  $^{b}$ Physics Division, National Technical University of Athens, \\15780 Zografou Campus, Athens, Greece}}\\

\vspace{.3cm}
{\normalsize { \it $^{c}$ Department of Theoretical Physics and Center for Astroparticle Physics (CAP)\\ 24 quai E. Ansermet, CH-1211 Geneva 4, Switzerland}}
\\[1.8in]

%%%%%%%%%%%%%%%%%%%%%%%%%%%%%%%%%%%%%%%%%%%%%%

{\bf ABSTRACT}\\[.01in]              %%%% Abstract - Pacs code - Keywords %%%%

\end{center}

\begin{quotation}
{In the framework of linearized non-minimal supergravity (20/20), we present the embedding of the 
$R+R^2$ model and we analyze its field spectrum. 
As usual, the auxiliary fields of the Einstein theory now become propagating, giving rise to additional degrees of freedom, which organize themselves into on-shell irreducible supermultiplets. 
By performing the analysis both in component and superspace formulations we identify the new supermultiplets. 
On top of the two massive chiral superfields reminiscent of the old-minimal supergravity embedding, 
the spectrum contains also a consistent physical, massive, vector supermultiplet 
and a tachyonic ghost, massive, vector supermultiplet.}

%${~~~}$ \newline
%PACS: 04.65.+e
%
%%%%%%%%%%%%%%%%%%%%%%%%%%%%%%%%%%%%%%%%%%%%%%
%
\endtitle                           %%%% End title page %%%%

\section{Introduction}

Supergravity, \cite{SG}, 
as the low energy limit of superstring theory, 
offers the proper setup to study  high energy gravitational phenomena. Among others, it provides an appropriate framework for the accommodation of cosmic inflation.  The constraints on the latter released by the Planck collaboration \cite{Ade:2013uln} 
favor inflationary models\hyphenation{mo-dels} which are characterized by plateau potentials with a tiny tensor-to-scalar ratio $r$ \cite{Lyth:1998xn}. 
Among the candidates is a higher curvature gravitational model, 
the Starobinsky model of inflation \cite{Starobinsky:1980te}
\be
\sqrt{-g}^{-1} {\cal L} = \frac12  M_P^2 R + \frac{M_P^2}{m^2} R^2~~, 
\ee
which stands out for its simplicity in providing a microscopic description of the mechanism responsible for the quasi de Sitter phase during inflation. 
This is a particular higher curvature gravitational theory of the type described in \cite{Stelle}.  It  is classically\hyphenation{cla-ssi-ca-lly} equivalent to a 
theory of standard gravitation coupled to an additional propagating real scalar degree of freedom \cite{Whitt:1984pd}, with a sufficiently flat potential at large values, ideal to drive inflation.

%Our objective is to construct an $R+R^2$ higher curvature supergravity theory.   
%To built higher curvature invariants one has to utilize the superfields of the gravitational supermultiplet. 
However, it is a well known fact that $4D,~\mathcal{N}=1$ supergravity does not have a unique off-shell description. There are two popular minimal formulations with 12 bosonic and 12 fermionic off-shell degrees of freedom (12/12):  the old-minimal \cite{Wess:1978bu} and the new-minimal \cite{west,Howe:1981et} supergravity.  In addition, there exists another one with 20 bosonic and 20 fermionic off-shell degrees of freedom, which still fill an irreducible supersymmetry multiplet, the 20/20  non-minimal  supergravity \cite{Siegel:1978mj}. 
%This is not an alien concept in supersymmetric theories; 
%the distribution  of the auxiliary degrees of freedom needed to complete a multiplet is not unique. 
%For the gravitational supermultiplet there exist the minimal 12/12 multiplets which contain 
%12 bosonic and 12 fermionic degrees of freedom. 
The Starobinsky model has been embedded in the old-minimal formulation  \cite{Ferrara:1978rk,Cecotti:1987sa,Kallosh:2013lkr,Farakos:2013cqa,Dalianis:2014aya} as well as the 
new-minimal formulation 
\cite{Farakos:2013cqa,Cecotti:1987qe,Cecotti:1987qr,Ferrara:2013rsa,Ferrara:2014cca} 
along with various modifications 
\cite{Ellis:2013xoa,Ellis:2013nxa,Ferrara:2013eqa,Ferrara:2014rya,Antoniadis:2014oya,Farakos:2014gba,Ellis:2014gxa,
Ketov:2014qha,Ferrara:2014yna,Ferrara:2014kva,Ketov:2014hya,Dall'Agata:2014oka,Terada:2014uia}. 
It has also been studied in the framework of gravitino condensation 
\cite{Alexandre:2013nqa,Alexandre:2014lla}. Nevertheless there is no analogue discussion for the non-minimal formulation of supergravity. The purpose of this work is exactly that: to demonstrate the construction of the $R+R^2$ Starobinsky model in the framework of non-minimal supergravity. For completeness, we would like to comment that there exist another non-minimal formulation
\cite{Lang:1985xk,Hayashi:1985vd,Aulakh:1985dn,Siegel:1986sv} with 16/16 degrees of freedom. However it is not an irreducible representation and can be decomposed to old-minimal supergravity with a chiral supermultiplet.

To outline the procedure, we start with the free theory of nonminimal supergravity which includes a set of dynamical components that describe gravity (helicity $\pm$ 2) with its superpartener, the gravitino (helicity $\pm$ 3/2) and another set of auxiliary components just so the SUSY algerbra will close off-shell. Afterwards we introduce the higher curvature terms of the form $R^2$. 
Due to the higher derivatives, the auxiliary fields of the free theory start propagating and organize themselves into supermultiplets. Nevertheless, these supermultiplets will have to be on-shell because only their dynamical degrees of freedom appear in the action, no auxiliary fields. The goal is to uncover these newly formed on-shell supermultiplets and their properties. In order to do that, we quickly realize that, we do not need to start with the full theory but its linearized version will do. The results of this analysis for the case of old-minimal supergravity \cite{Ferrara:1978rk,Cecotti:1987sa} are two physical, massive, chiral supermultiplets and for the case of new-minimal supergravity \cite{Cecotti:1987qe} is a physical, massive, vector supermultiplet.     
%
%The superspace embeddings exhibit differences between them which originate from the different underlying superspace structure. 
%To uncover the multiplet properties of these higher curvature supergravities it is sufficient to turn to the linearized level, 
%before turning to the full theory. 
%Indeed, an analysis of the linearized theories shows that 
%the new-minimal $R+R^2$ supergravity is equivalent to 
%{\it standard supergravity} coupled to a massive vector multiplet \cite{Cecotti:1987qe}, 
%while the old-minimal $R+R^2$ supergravity is equivalent to two chiral superfields coupled to 
%{\it standard supergravity} .   
%This equivalence also holds for the full theories.  
%and the 20/20 %\cite{Breitenlohner:1976nv,Breitenlohner:1977jn,Girardi:1984eq,Garreis:1987cv,Grimm:1984pj} 
%supergravity multiplets. 
%In this work we focus on the 20/20 supergravity and we study  
%the equivalent theory for the $R + R^2$ model in this formulation.  
%A full treatment of the theory would be welcome, 
%but to uncover the multiplet structure we only need the theory at the linearized level. 

Linearized supergravity is nothing else but the theory of massless, irreducible representation\hyphenation{re-pre-se-nta-tion} of the super-Poincar\'{e} group with superhelicity $Y$=$3/2$. The superspace and component\hyphenation{co-mpo-nent} formulation of the massless, arbitrary superhelicity, irreducible representations and their properties have been studied in detail in a series of papers \cite{Gates:2014tea,Kuzenko:1993jq,Kuzenko:1993jp}.
For our purpose, we will use the formalism and the results of \cite{Gates:2014tea} and adapt them for the case of superhelicity $Y$=$3/2$. 

The presentation of this work is organized in the following way. In Section \ref{S1} we briefly review the results of \cite{Gates:2014tea} for the case of linearized, non-minimal supergravity in both superspace and components. Then in section \ref{S2} we construct the $R^2$ action in superspace and project to components. In section \ref{S3} we combine the two previous results to construct the Starobinsky model ($R+R^2$) in this framework and study its spectrum. We perform the analysis\hyphenation{a-na-ly-sis} at the component level for both bosons and fermions. At the end we verify our results by performing a duality in
superspace that reveals exactly the same spectrum and demonstrates\hyphenation{de-mon-stra-tes} the classical equivalence between the $R+R^2$ theory of non-minimal supergravity and the {\it standard} non-minimal supergravity coupled to two massive chiral supermultiplets, a massive vector supermultiplet and a tachyonic ghost, massive vector supermultiplet. An interesting remark is that all of the massive supermultiplets turn out to have equal masses. 
%
%
%
%which is a formalism applicable to higher spin supersymmetric theories in general. 
%This setup relies on a compensator superfield which 
%guarantees the gauge invariance and also provides the additional 
%auxiliary component fields. 
%As usual the gauge superfields of the gravitational sector contain various linearized geometrical objects. 
%After identifying these superfields which also serve as building blocks we construct the higher curvature terms 
%and study the $R+R^2$ theory. 
%Finally, as an independent verification of our findings we illustrate the 
%classical equivalence in superspace of the higher curvature theory to the {\it standard} 20/20 
%supergravity coupled to matter. 

\section{ \fontsize{14}{8} \selectfont \!\!\!\!\!\!\!Superhelicity $Y$=$\frac{3}{2}$ as  Linearized Non-Minimal Supergravity}
\label{S1}

From the investigation of free, massless, higher superspin theories \cite{Gates:2014tea} we can extract the  $4D, \mathcal{N}=1$ superspace action for linearized non-minimal supergravity     
\bea{ll}
\label{AAA}
\mathcal{S}_{R}=\int d^8z&\left\{\vphantom{\frac12} ~~~H^{\a\ad}\D^{\g}\Dd^2
\D_{\g}H_{\a\ad}\right.\\
&~~-2~H^{\a\ad}\Dd_{\ad}\D^2\chi_{\a}+c.c.\\
&~~-2~\chi^{\a}\D^2\chi_{\a}+c.c.\IEEEyesnumber\\
&~~+\left.\vphantom{\frac12}2~\chi^{\a}\D_{\a}
\Dd^{\ad}\bar{\chi}_{\ad}\right\},\\
\eea
which contains the real bosonic superfield $H_{\a\ad}$ and the fermionic superfield $\chi_{\a}$ as a compensator\hyphenation{co-mpe-nsa-tor}. The action is invariant under the following transformation
\bea{ll}
\IEEEyesnumber
\delta_G H_{\a\ad}&=\D_{\a}\bar{L}_{\ad}-\Dd_{\ad}L_{\a},\IEEEyessubnumber\\
\delta_G \chi_{\a}&=\Dd^2L_{\a}+\D^{\b}\Lambda_{\a\b},\IEEEyessubnumber
\eea
which forces the following Bianchi Identities
\bea{l}
\IEEEyesnumber
\label{B.I}
\Dd^{\ad}T_{\a\ad}-\Dd^2 G_{\a}=0,
\IEEEyessubnumber\label{B.I.1}\\
\frac{1}{2!}\D_{(\a}G_{\b)}=0.\IEEEyessubnumber\label{B.I.2}
\eea
The superfields $T_{\a\ad}$ and $G_{\a}$ are the variations of the action (\ref{AAA}) 
with respect to the unconstrained superfields $H_{\a\ad}$ and $\chi_{\a}$. Their explicit expressions are 
\bea{ll}
\IEEEyesnumber
\label{E.Q.I}
T_{\a\ad}&=2\D^\g\Dd^2\D_\g H_{\a\ad}+2\left(\D_{\a}\Dd^2\bar{\chi}_{\ad}-
\Dd_{\ad}\D^2\chi_{\a}\right),\IEEEyessubnumber\\
G_{\a}&=-2\D^2\Dd^{\ad}H_{\a\ad}-4\D^2\chi_{\a}+2\D_{\a}\Dd^{\ad}\bar{\chi}_{\ad}. \IEEEyessubnumber
\eea
The two superfields $T_{\a\ad}$ and  $G_{\a}$ in (\ref{E.Q.I}) 
have mass dimensionality 
$\left[T_{\a\ad}\right]=2$,~$\left[G_{\a}\right]=3/2$\footnote{The highest spin component of $H_{\a\ad}$ is a propagating boson.}.

To prove that indeed this 
action describes the desired representation, using the equations 
of motion we can now show that a gauge invariant chiral superfield $F_{\a\b\g}$ exists ([$F_{\a\b\g}$]=$5/2$)
\bea{l}
F_{\a\b\g}=\frac{1}{3!}\Dd^2\D_{(\a}\pa_{\b}{}^{\ad}H_{\g)\ad},\IEEEyesnumber
\eea
and on-shell ($T_{\a\ad}$ =  $G_{\a}$ = 0), it satisfies the desired constraints in order to describe a 
super-helicity $Y$=$3/2$ system
\bea{l}
\Dd_{\ad}F_{\a\b\g}=0~,~\D^{\a}F_{\a\b\g}=0.\IEEEyesnumber
\eea

At the component level, the above superspace action describes the dynamics of  the following bosons
\bea{l}
u_{\a\ad}\equiv\frac{1}{2}\left\{\D_{\a}\bar{G}_{\ad}-\Dd_{\ad}G_{\a}\right\}|,~
v_{\a\ad}\equiv -\frac{i}{2}\left\{\D_{\a}\bar{G}_{\ad}+\Dd_{\ad}G_{\a}\right\}|,\\
S\equiv\frac{1}{2}\left\{\D^{\a}G_{\a}+\Dd^{\ad}\bar{G}_{\ad}\right\}|,~~~
P\equiv -\frac{i}{2}\left\{\D^{\a}G_{\a}-\Dd^{\ad}\bar{G}_{\ad}\right\}|,\IEEEyesnumber\label{Comp.}\\
A_{\a\ad}\equiv T_{\a\ad}|+\frac{1}{3}\left(\D_{\a}\bar{G}_{\ad}-\Dd_{\ad}G_{\a}\right)|,\\
h_{\a\b\ad\bd}\equiv\frac{1}{2(2!)^2}\left[\D_{(\a},\Dd_{(\ad}\right]H_{\b)\bd)}|,\\
h\equiv\frac{1}{8}\left[\D^{\a},\Dd^{\ad}\right]H_{\a\ad}|+\frac{1}{2}\left(\D^{\a}\chi_{\a}+\Dd^{\ad}\bar{\chi}_{\ad}\right)|,
\eea
namely, in 4-component notation, of three vectors $A_\mu~(A_{\alpha\dot \alpha})$, $u_\mu~(u_{\alpha\dot \alpha})$ and $v_\mu~(v_{\alpha\dot \alpha})$, three scalars $(S,P,h)$ and a symmetric traceless rank-2 tensor (the graviton) $h_{\mu\nu}~(h_{\alpha\beta\dot \alpha\dot \beta})$. 
The corresponding gauge transformations acting on the bosons are 
\bea{l}
\d_G A_{\a\ad}=0~,~\d_G u_{\a\ad}=0~,~\d_G v_{\a\ad}=0,\\
\d_G S=0~,~\d_G P=0,\IEEEyesnumber\\
\d_G h_{\a\b\ad\bd}=\frac{1}{(2!)^2}\pa_{(\a(\ad}\zeta_{\b)\bd)},\\
\d_G h=\frac{1}{4}\pa^{\a\ad}\zeta_{\a\ad}~,~~\zeta_{\a\ad}=\frac{i}{2}\left(\D_{\a}\bar{L}_{\ad}+\Dd_{\ad}L_{\a}\right)| ,
%\\
\eea
which leave 4 degrees of freedom for each vector, 1 for each scalar and 5 for the symmetric traceless tensor, a total of 20 degrees of freedom to fill the bosonic part of the non-minimal irreducible supersymmetric multiplet. 
The bosonic sector of the  Lagrangian density is
\bea{ll}
\mathcal{L}_{R}|_{{}_B}=
\mathcal{L}_{h=\pm 2} +\frac{1}{6}~u^{\a\ad}u_{\a\ad}-\frac{1}{2}~v^{\a\ad}v_{\a\ad}
+\frac{3}{16}~A^{\a\ad}A_{\a\ad}-\frac{1}{8}~SS-\frac{1}{8}~PP,\IEEEyesnumber
\eea
where $\mathcal{L}_{h=\pm 2}$ describes a massless helicity $\pm 2$ particle
\bea{ll}
\mathcal{L}_{h=\pm 2}
&=h^{\a\b\ad\bd}\Box h_{\a\b\ad\bd}-h^{\a\b\ad\bd}\pa_{\a\ad}\pa^{\g\gd}h_{\g\b\gd\bd}
+2~h^{\a\b\ad\bd}\pa_{\a\ad}\pa_{\b\bd}h-6~h\Box h,\IEEEyesnumber \\
&=\frac{1}{\kappa^2}[\sqrt{-g}R]|_\text{linearized},
\eea
and $[\sqrt{-g}R]|_\text{linearized}$ is the linearized  Einstein-Hilbert Lagrangian, keeping only the terms  
quadratic in the fields. At this linear approximation, the Ricci scalar takes the form (up to an overall normalization)
\be
R=\pa^{\a\ad}\pa^{\b\bd}h_{\a\b\ad\bd}-6\Box h,
\ee
and its mass dimension is $\left[R\right]=3$. 
The Ricci scalar is part of the completely antisymmetric  $\theta\bar{\theta}$ term in the expansion of the $T_{\a\ad}$ superfield, 
specifically 
\be
\left[\D^{\a},\Dd^{\ad}\right]T_{\a\ad}|=-4R-6\pa^{\a\ad}v_{\a\ad}. \IEEEyesnumber
\ee
Also, the linearized Ricci tensor is 
\be
R_{\a\b\ad\bd}=\Box h_{\a\b\ad\bd}-\frac{1}{2!2!}\pa_{(\a(\ad}\pa^{\g\gd}h_{\g\b)\gd\bd)}+\frac{1}{2!2!}\pa_{(\a(\ad}\pa_{\b)\bd)}h,
\ee
and it resides in the fully symmetric part of the $\theta\bar{\theta}$ term of $T_{\a\ad}$ 
\be
\frac{1}{2!2!}\left[\D_{(\a},\Dd_{(\ad}\right]T_{\b)\bd)}|=\frac{2}{2!2!}\pa_{(\a(\ad}v_{\b)\bd)}-4 R_{\a\b\ad\bd},\label{R.T}
\ee
while it satisfies
\be
\pa^{\b\bd}R_{\a\b\ad\bd}+\frac{1}{4}\pa_{\a\ad}R=0 . 
\ee

Similarly for the fermionic sector, we have the following components 
\bea{l}
\b_{\a}\equiv -\frac{1}{4}\left\{\D_{\a}\Dd^{\ad}\bar{G}_{\ad}-i\pa_{\a}{}^{\ad}\bar{G}_{\ad}\right\}|~~,\\
\rho_{\a}\equiv G_{\a}|~~,\\
\psi_{\a\b\ad}\equiv\frac{\sqrt{2}}{2!}\Dd^2\D_{(\a}H_{\b)\ad}|~~,\IEEEyesnumber\label{Comp.2}\\
\psi_{\a}\equiv -\sqrt{2}\left\{\D^2\D^{\ad}H_{\a\ad}+2\D^2\chi_{\a}\right\}|.
\eea
The gauge transformations of the fermionic  fields are 
\bea{ll}
\d_G\rho_{\a}=0~,~&\d_G\psi_{\a\b\ad}=\frac{1}{2!}\pa_{(\a\ad}\xi_{\b)},\\
\d_G\b_{\a}=0~,~&\d_G\psi_{\a}=-\pa_{\a}{}^{\ad}\bar{\xi}_{\ad},\IEEEyesnumber
\eea
with $\xi_{\a}=-i\sqrt{2}~\Dd^2L_{\a}|$.
The corresponding free Lagrangian is
\bea{ll}
\mathcal{L}_{R}|_{{}_F}&=\mathcal{L}_{h=\pm 3/2}~+\b^{\a}\r_{\a}~+\bar{\b}^{\ad}\bar{\r}_{\ad},\IEEEyesnumber
\eea
where $\mathcal{L}_{h=\pm 3/2}$ describes a massless Rarita-Swinger field (gravitino with helicity $\pm$ $3/2$) 
\be
\mathcal{L}_{h=\pm 3/2}= ~i\bar{\psi}^{\a\ad\bd}\pa^{\b}{}_{\bd}\psi_{\a\b\ad} 
-\frac{3}{4}i\bar{\psi}^{\ad}\pa^{\a}{}_{\ad}\psi_{\a} 
+ \left( \frac{i}{2}\psi^{\a\b\ad}\pa_{\b\ad}\psi_{\a}~+c.c. \right) . 
\ee
The linearized fermionic curvatures are 
\bea{ll}
\IEEEyesnumber
&R_{\a}=i\sqrt{2}\pa^{\b\bd}\psi_{\a\b\bd}+\frac{3i}{\sqrt{2}}\pa_{\a}{}^{\ad}\bar{\psi}_{\ad},\IEEEyessubnumber\\
&R_{\a\b\ad}=\frac{i\sqrt{2}}{2!}\pa_{(\a}{}^{\bd}\bar{\psi}_{\b)\ad\bd}+\frac{i}{\sqrt{2}2!}\pa_{(\a\ad}\psi_{\b)},\IEEEyessubnumber
\eea
and they are the (anti)symmetric part of the $\bar{\theta}$ term of superfield $T_{\a\ad}$
\bea{ll}
\IEEEyesnumber
\frac{1}{2!}\Dd_{(\ad}T_{\a\bd)}|=\bar{R}_{\a\ad\bd},\IEEEyessubnumber\\
{}\\
\Dd^{\ad}T_{\a\ad}|=R_{\a}-4\b_{\a}-i\pa_{\a}{}^{\ad}\r_{\ad}\IEEEyessubnumber.
\eea
Finally they satisfy 
\be
\pa^{\a\ad}\bar{R}_{\a\ad\bd}-\frac{1}{2}\pa^{\b}{}_{\bd}R_{\b}=0.
\ee
%%%%%%%%%%%%%%%%%%%%%%%%%%%%%%%%%%%%%%%%%%%%%%

\section{Constructing the $\mathbf R^2$ theory}
\label{S2}
~~~~Now we turn to the construction of a gauge invariant, higher derivative superspace action, such that it will generate $R^2$ terms. The reason that we restrict ourselves only to $R^2$ terms and we do not include  for example the square of the Ricci tensor, 
or equivalently the Weyl tensor square, is that the inclusion of the latter terms will lead to ghost and/or tachyons states in the spectrum \cite{Stelle,Ferrara:1978rk} .

To proceed in our cosntruction, we recall that the available gauge invariant objects are the superfields  $T_{\a\ad}$,~$G_{\a}$ and $F_{\a\b\g}$. However $F_{\a\b\g}$, due to its chiral property and its index structure, it can only couple to itself, giving a term of the form $F^{\a\b\g}F_{\a\b\g}$. But such an object will give rise to the square of the Weyl tensor, so it is rejected. The rest of the objects could be combined in many different ways. We organize them in the following manner. 

The general structure of all possible terms that we are interested in, are schematically of the form
\be
T^n~D^k~G^l,
\ee 
which means that any possible term will include $n$ $T_{\a\ad}$'s, $k$ superspace covariant derivatives and $l$ $G_{\a}$'s. The dimensionality of these terms is 
\be
2n+\frac{k}{2}+\frac{3l}{2}. 
\ee
Then, if we project to components, we have to integrate over superspace
$\Dd^2\D^2\left(T^n~D^k~G^l\right)|$, and therefore the mass dimension of the component terms that we can, in principle, construct is
\be
2n+\frac{k}{2}+\frac{3l}{2}+2~.
\ee
The finall step is the fact that the desired $R^2$ term has dimensionality 6 and  we require to have expressions quadratic in the components (linear approximation). Therefore we must have  
\bea{cc}
\IEEEyesnumber
2n+\frac{k}{2}+\frac{3l}{2}+2=6,\IEEEyessubnumber\\
n+l=2. \IEEEyessubnumber
\eea 
The solutions of this Diophantine system, and the corresponding terms allowed are given 
in the following table 
\begin{center}
\begin{tabular}{|c|c|c|c|}
 \hline 
 $n$ & $k$ & $l$ & term \\  
 \hline 
 2 & 0 & 0 & $T^{\a\ad}T_{\a\ad}$ \\ 
 \hline 
 1 & 1 & 1 & $T^{\a\ad}\Dd_{\ad}G_{\a}+c.c.$ \\ 
 \hline 
 \multirow{3}{*}{0} & \multirow{3}{*}{2} & \multirow{3}{*}{2} & $G^{\a}\D_{\a}\Dd^{\ad}\bar{G}_{\ad}$ \\ 
 \cline{4-4} 
  &  &  & $G^{\a}\Dd^{\ad}\D_{\a}\bar{G}_{\ad}$ \\ 
 \cline{4-4} 
  &  &  & $G^{\a}\Dd^2 G_{\a}+c.c.$ \\ 
 \hline 
 \end{tabular}  
\end{center}
Note that we have not included the term $G^{\a}\D^2 G_{\a}+c.c.$ since it is zero due to (\ref{B.I.2}). 
Moreover because of equation (\ref{B.I.1}) 
the terms $T^{\a\ad}\Dd_{\ad}G_{\a}+c.c.$~and~$G^{\a}\Dd^2 G_{\a}+c.c.$ are identical.

Hence the $R^2$ superspace action must be of the form
\be
S_{R^2}=\int d^8z\left\{g_0 T^{\a\ad}T_{\a\ad}+g_1 G^{\a}\D_{\a}\Dd^{\ad}\bar{G}_{\ad}+g_2 G^{\a}\Dd^{\ad}\D_{\a}\bar{G}_{\ad}+(g_3 G^{\a}\Dd^2 G_{\a}+c.c.) \right\}\label{S.R2}
\ee
where $g_0~,~g_1~,g_2~\in \mathbb{R}$. 
Now what remains is to project this action to components and pick the coefficients in a way such 
that we generate  $R^2$ terms and 
canonical kinematic terms for any additional propagating fields.  
%diagonal (no interactions) kinetic energy terms for the so far auxiliary terms. 

%%%%%%%%%%%%%%%%%%%%%%%%%%%%%%%%%%%%%%%%%%%%%%

%\subsection{Component Projection: Bosons}

The component Lagrangian we get from the above action (\ref{S.R2}) is
\bea{ll}
\mathcal{L}_{R^2}=&g_0 \Dd^2\D^2(T^{\a\ad}T_{\a\ad})|+g_1 \Dd^2\D^2(G^{\a}\D_{\a}\Dd^{\ad}\bar{G}_{\ad})|\IEEEyesnumber\\
&+g_2 \Dd^2\D^2(G^{\a}\Dd^{\ad}\D_{\a}\bar{G}_{\ad})|+\left\{g_3 \Dd^2\D^2(G^{\a}\Dd^2 G_{\a})|+c.c.\right\}.
\eea
The basic rules for projection are 
\begin{enumerate}
\item Use the `Leibniz' rule
\bea{ll}
\Dd^2\D^2(AB)|=&\Dd^2\D^2A|B|+(-1)^{\e(A)}\Dd^{\rd}\D^2A|\Dd_{\rd}B|+\D^2A|\Dd^2B|\\
&+(-1)^{\e(A)}\Dd^2\D^{\r}A|\D_{\r}B|-\Dd^{\rd}\D^{\r}A|\Dd_{\rd}\D_{\r}B|+(-1)^{\e(A)}\D^{\r}A|\Dd^2\D_{\r}B|\\
&+\Dd^2A|\D^2B|+(-1)^{\e(A)}\Dd^{\rd}A|\Dd_{\rd}\D^2B|+A|\Dd^2\D^2B|,\IEEEyesnumber
\eea
where $\e$ is zero for bosonic and one for fermionic superfields. 
\item Use the Bianchi identities (\ref{B.I}).
\item Use the component definitions of (\ref{Comp.}) and (\ref{Comp.2}). 
\end{enumerate}

First we focus on the bosonic sector of the theory, 
therefore we restrict the above calculation to the bosonic part of the projection. 
That means, we keep only the terms with even number of $\D$'s when acting on a bosonic superfield (like $T_{\a\ad}$) 
and with odd number of $\D$'s when acting on a fermionic superfield (like $G_{\a}$). 
We get 
\bea{ll}
\IEEEyesnumber
\mathcal{L}_{R^2}|_{{}_B}=I_0|_{{}_B}+I_1|_{{}_B}+I_2|_{{}_B}+I_3|_{{}_B},
\eea
with
\bea{ll}
\IEEEyesnumber
I_0|_{{}_B}=g_0 \Dd^2\D^2(T^{\a\ad}T_{\a\ad})|_{{}_B},\IEEEyessubnumber\\
I_1|_{{}_B}=g_1 \Dd^2\D^2(G^{\a}\D_{\a}\Dd^{\ad}\bar{G}_{\ad})|_{{}_B},\IEEEyessubnumber\\
I_2|_{{}_B}=g_2 \Dd^2\D^2(G^{\a}\Dd^{\ad}\D_{\a}\bar{G}_{\ad})|_{{}_B},\IEEEyessubnumber\\
I_3|_{{}_B}=g_3 \Dd^2\D^2(G^{\a}\Dd^2 G_{\a})|_{{}_B}+c.c.\IEEEyessubnumber
\eea
It is evident that $I_0|_{{}_B}$ includes a term proportional to $\left[\D^{(\r},\Dd^{(\rd}\right]T^{\a)\ad)}|\left[\D_{(\r},\Dd_{(\rd}\right]T_{\a)\ad)}$, which based on (\ref{R.T})
makes it obvious that it generates the Ricci tensor square, $R^{\a\b\ad\bd}R_{\a\b\ad\bd}$, a term that is not considered here (as it leads to ghost and/or tachyonic states \cite{Stelle,Ferrara:1978rk}).  On top of that, such a term can not be canceled by any of the other contributions to the bosonic Lagrangian. Therefore the only possibility out of that, is to choose
\be
g_0=0.
\ee

The rest of the terms are relevant  
and after putting everything together, we find that the total bosonic sector is 
\bea{ll}
\mathcal{L}_{R^2}|_{{}_B}=&~~\frac{1}{4}\[g_1-g_2-g^R_3\]~A^{\a\ad}\pa_{\a\ad}\pa^{\b\bd}A_{\b\bd}\\
&+\frac{1}{6}\[4g_1-g_2+2g^R_3\]~A^{\a\ad}\pa_{\a\ad}\pa^{\b\bd}u_{\b\bd}\\
&+\frac{1}{9}\[4g_1-7g_2+8g^R_3\]~u^{\a\ad}\pa_{\a\ad}\pa^{\b\bd}u_{\b\bd}~+\[g_2-2g^R_3\]~u^{\a\ad}\Box u_{\a\ad}\\
&+\[4g_1-7g_2+8g^R_3\]~v^{\a\ad}\pa_{\a\ad}\pa^{\b\bd}v_{\b\bd}~+\[g_2+2g^R_3\]~v^{\a\ad}\Box v_{\a\ad}\IEEEyesnumber\label{B.R2}\\
&+2\[4g_1-5g_2+6g^R_3\]~v^{\a\ad}\pa_{\a\ad}R  ~~+\[3g^I_3\]~A^{\a\ad}\pa_{\a\ad}\pa^{\b\bd}v_{\b\bd}\\
&+\[-g_1+\frac{1}{2}g_2\]~S\Box S
%~~~~~~~~~~~~~~~~~~~~
+\[2g^I_3\]~A^{\a\ad}\pa_{\a\ad}R\\     
&+\[-g_1+\frac{1}{2}g_2\]~P\Box P
%~~~~~~~~~~~~~~~~~~~~
+\[-4g^I_3\]~u^{\a\ad}\Box v_{\a\ad}\\
&-4\[g_1-g_2+g^R_3\]R^2
%~~~~~~~~~~~~~~~~
+\[-\frac{4}{3}g^I_3\]~u^{\a\ad}\pa_{\a\ad}R, 
\eea
where $g^R_3$and $g^I_3$ are the real and imaginary parts of $g_3$. 
Notice that the higher curvature terms are accompanied by kinematic terms for all the previously auxiliary fields. 
This is a standard property of higher curvature supergravity. 

%%%%%%%%%%%%%%%%%%%%%%%%%%%%%%%%%%%%%%%%%%%%%%

Similarly, we find that  the fermionic sector is 
\bea{ll}
\mathcal{L}_{R^2}|_{{}_F}=&-4\[4g_1-7g_2+8g^R_3\]~i\bar{\b}^{\ad}\pa^{\a}{}_{\ad}\b_{\a} \\
&-\frac{1}{4}\[4g_1-7g_2+8g^R_3\]~i\bar{\r}^{\ad}\Box\pa^{\a}{}_{\ad}\r_{\a} \\
&+\[4g_1+3g_2-8g^R_3\]~\b^{\a}\Box\r_{\a}+c.c.\IEEEyesnumber\label{F.R2}\\
&+\[g_2\]~i\bar{R}^{\ad}\pa^{\a}{}_{\ad} R_{\a}\\
&-4\[g_2-g_3\]~i\bar{\b}^{\ad}\pa^{\a}{}_{\ad}R_{\a}+c.c.\\
&-\[g_2-g_3\]~\r^{\a}\Box R_{\a}+c.c.
\eea

%%%%%%%%%%%%%%%%%%%%%%%%%%%%%%%%%%%%%%%%%%%%%%

\section{The spectrum of  $\mathbf{R+R^2}$ non-minimal supergravity}
\label{S3}
So far we have developed the superspace action for the $R$ and $R^2$ theories. In this section we combine them in order to study the spectrum of the $R+R^2$ theory. Specifically we will analyze the propagating degrees of freedom of the Lagrangian 
\be
\mathcal{L}=\mathcal{L}_{R}+\frac{1}{m^2}\mathcal{L}_{R^2}~. 
\ee
To do this we must first bring the full Lagrangian 
into a diagonal form and subsequently study their field equations. 
Typically one can achieve that, by doing redefinitions of the various fields and a clever choice of coefficients. But, in this case due to the fact that the $\mathcal{L}_{R}$ is already diagonal, we can not perform any redefinitions and the only thing left to do is to choose appropriately the  coefficients of the  non-diagonal terms. 

\subsection{Bosonic sector spectrum}
Following the previously explained strategy,
we must impose the constraints 
\be
 \left. \begin{array}{ll}
         &4g_1-g_2+2g^R_3=0\\
       &4g_1-5g_2+6g^R_3=0\\
       &g^I_3=0  \end{array} \right\} {\ } 
g_1=-\frac{1}{4}g~,~g_2=g_3=g~,~g\in \mathbb{R}.\IEEEyesnumber\label{Coef}
\ee
With the above coefficients (\ref{Coef}), we find that the linearized, bosonic part of the component Lagrangian  is 
\bea{ll}
\mathcal{L}|_{{}_{B}}=&~\mathcal{L}_{h=\pm 2} +\frac{g}{m^2}R^2\\
&+\frac{3}{16}~A^{\a\ad}A_{\a\ad} -\frac{9}{16}\frac{g}{m^2}~A^{\a\ad}\pa_{\a\ad}\pa^{\b\bd}A_{\b\bd}\\
&+\frac{1}{6}~u^{\a\ad}u_{\a\ad} -\frac{g}{m^2}~u^{\a\ad}\Box u_{\a\ad}\\
&-\frac{1}{2}~v^{\a\ad}v_{\a\ad} +3\frac{g}{m^2}~v^{\a\ad}\Box v_{\a\ad}\IEEEyesnumber\\
&-\frac{1}{8}~S^2 +\frac{3}{4}\frac{g}{m^2}~S\Box S\\
&-\frac{1}{8}~P^2 +\frac{3}{4}\frac{g}{m^2}~P\Box P . 
\eea
The equations of motion for the various fields and the degrees of freedom they allow to propagate are: 
\begin{enumerate}
\item For $A_{\a\ad}$ we have
\be
A_{\a\ad}-3\frac{g}{m^2}\pa_{\a\ad}\pa^{\b\bd}A_{\b\bd}=0~~\rightsquigarrow~~\Box\pa^{\a\ad}A_{\a\ad}=\frac{m^2}{6g}\pa^{\a\ad}A_{\a\ad}.
\ee
From the left equation we see that three of the degrees of freedom of the vector field $A_{\a\ad}$ remain auxiliary and are solved in terms of the scalar $\pa^{\a\ad}A_{\a\ad}$.
From the right equation we see that for 
$g>0$, $\pa^{\a\ad}A_{\a\ad}$ is a physical, real, propagating, massive scalar with mass $\mu^2$=$m^2/6g$. 

\item For $u_{\a\ad}$ we find
\be
\frac{1}{6}u_{\a\ad}-\frac{g}{m^2}\Box u_{\a\ad}=0~~\rightsquigarrow~~\Box u_{\a\ad}=\frac{m^2}{6g}u_{\a\ad}.
\ee
This describes the propagation of a real, massive, scalar $\pa^{\a\ad}u_{\a\ad}$ with equations of motion 
$\Box\pa^{\a\ad}u_{\a\ad}$=$\frac{m^2}{6g}\pa^{\a\ad}u_{\a\ad}$ and  mass $\mu^2$=$m^2/6g$, 
and the propagation of a real, massive vector with the same mass described by the divergent-less field defined as  $\hat{u}_{\a\ad}$=$u_{\a\ad}-\frac{3g}{m^2}\pa_{\a\ad}\pa^{\b\bd}u_{\b\bd}$, with equations of motion 
~~$\Box \hat{u}_{\a\ad}$=$\frac{m^2}{6g}\hat{u}_{\a\ad}$. 
Both of them are tachyonic ghosts (for $g>0$) since they appear in the Lagrangian with an opposite overall sign.
\item For $v_{\a\ad}$ we have 
\be
-\frac{1}{2}v_{\a\ad}+3\frac{g}{m^2}\Box v_{\a\ad}=0~~\rightsquigarrow~~\Box v_{\a\ad}=\frac{m^2}{6g}v_{\a\ad}.
\ee
As before this equation includes both the spin zero part, described by $\pa^{\a\ad}v_{\a\ad}$ and the spin one part , described by the $\hat{v}_{\a\ad}$=$v_{\a\ad}-\frac{3g}{m^2}\pa_{\a\ad}\pa^{\b\bd}v_{\b\bd}$. Both of them have the same mass $\mu^2$=$m^2/6g$ and are physical for $g>0$.

\item For $S$ we find
\be
-\frac{1}{2}S+3\frac{g}{m^2}\Box S~~\rightsquigarrow~~\Box S=\frac{m^2}{6g}S,
\ee
which describes a physical ($g>0$), real, massive propagating scalar with mass $\mu^2$=$m^2/6g$.

\item For $P$ we find 
\be
-\frac{1}{2}P+3\frac{g}{m^2}\Box P~~\rightsquigarrow~~\Box P=\frac{m^2}{6g}P. 
\ee
Same as $S$, it describes a physical ($g>0$), real, massive propagating scalar with mass $\mu^2$=$m^2/6g$.

\item 
The gravitational sector of the action is
\bea{lll}
\mathcal{S'}|_{{}_{B}}=&~\int d^4x\mathcal{L}_{h=\pm 2}+\frac{g}{m^2}\int d^4x R^2,
\IEEEyesnumber
\eea
which can be re-expressed with the help of a Lagrange multiplier $\phi$ in the following form 
\bea{lll}
\mathcal{S'}|_{{}_{B}}=&~\int d^4x\mathcal{L}_{h=\pm 2}+f\int d^4x\phi R -\frac{f^2}{4g}m^2\int d^4x \phi^2,
\IEEEyesnumber
\eea
where $[\phi]=1$. 
Now we perform the following redefinition of $h$
\be
h\to h+c\phi.
\ee
The change of $\mathcal{L}_{h=\pm 2}$ is
\be
\delta\mathcal{L}_{h=\pm 2}=2c\phi R -6c^2\phi\Box\phi,
\ee
and the change of $R$ is
\be
\delta R=-6c\Box\phi.
\ee
Therefore we get for $\mathcal{S'}|_{{}_{B}}$
\bea{ll}
\mathcal{S'}|_{{}_{B}}=&\int d^4x\mathcal{L}_{h=\pm 2}+(2c+f)\int d^4x\phi R\IEEEyesnumber\\
&-6c(c+f)\int d^4x\phi\Box\phi ~-\frac{f^2}{4g}m^2\int d^4x \phi^2.
\eea
We choose $c$ such that the cross term vanish
\be
2c+f=0,
\ee 
hence we get
\bea{ll}
\mathcal{S'}|_{{}_{B}}=&\int d^4x\mathcal{L}_{h=\pm 2}\IEEEyesnumber\\
&+\frac{3}{2}f^2\int d^4x\phi\Box\phi ~-\frac{f^2}{4g}m^2\int d^4x \phi^2,
\eea
which describes a helicity $\pm 2$ and a physical (for $g>0$), real, massive, scalar $\phi$ with mass $\mu^2=m^2/(6g)$.
\end{enumerate}
To summarize, beside the helicity $\pm 2$ system,  
the spectrum organizes into two physical massive chiral supermultiplets $(\pa^{\a\ad}A_{\a\ad},\phi)$ and $(S,P)$, 
one physical massive vector supermultiplet $(\hat{v}_{\a\ad},\pa^{\a\ad}v_{\a\ad})$ 
and one tachyonic - ghost massive vector supermultiplet $(\hat{u}_{\a\ad},\pa^{\a\ad}u_{\a\ad})$. 

%%%%%%%%%%%%%%%%%%%%%%%%%%%%%%%%%%%%%%%%%%%%%%

\subsection{Fermionic sector spectrum}

In order to verify the fermionic spectrum, we start with equation (\ref{F.R2}) and make the same choice of coefficients as in (\ref{Coef}), which give
\bea{lll}
\label{ferm-lagr}
\mathcal{L}|_{{}_{F}}=&~\mathcal{L}_{h=\pm 3/2}+\b^{\a}\r_{\a}
-6\frac{g}{m^2}\b^{\a}\Box\r_{\a}\IEEEyesnumber\\
&+\bar{\b}^{\ad}\bar{\r}_{\ad}-6\frac{g}{m^2}\bar{\b}^{\ad}\Box\bar{\r}_{\ad} 
+i\frac{g}{m^2}\bar{R}^{\ad}\pa^{\a}{}_{\ad}R_{\a}.
\eea
The equations of motion for the various fields are
\begin{enumerate}

\item From  $\b_{\a}$ and $\bar{\b}_{\ad}$ we find
\be
\Box\r_{\a}=\frac{m^2}{6g}\r_{\a}~,~\Box\bar{\r}_{\ad}=\frac{m^2}{6g}\bar{\r}_{\ad},
\ee
which describe a pair of massive Weyl spinors with Dirac mass $\mu^2=m^2/(6g)$. 

\item From $\r_{\a}$ and $\bar{\r}_{\ad}$ we find
\be
\Box\b_{\a}=\frac{m^2}{6g}\b_{\a}~,~\Box\bar{\b}_{\ad}=\frac{m^2}{6g}\bar{\b}_{\ad},
\ee
which again describe a pair of massive Weyl spinors with Dirac mass $\mu^2=m^2/(6g)$. 
%\item For $\psi_{\a}$ and $\bar{\psi}_{\ad}$:
%\be
%\Box(NR)_{\a}=\frac{m^2}{6g}(NR)_{\a}~,~\Box(\bar{N}\bar{R})_{\ad}=\frac{m^2}{6g}(\bar{N}\bar{R})_{\ad}
%\ee 
Note that, in order to reveal the fermions that belong into the tachyonic - ghost vector multiplet, we have to diagonalize the Lagrangian (\ref{ferm-lagr}). Once we do that, we will get one positive and one negative eigenvalue, which signals the propagation 
of one physical and one tachyonic - ghost fermion.

\item The rest of the action includes $\mathcal{L}_{h=\pm 3/2}$ and $R_{\a}$ and can be expressed in the following way 
\bea{ll}
\mathcal{S'}|_{{}_{F}}=\int d^4x\mathcal{L}_{h=\pm 3/2}&+ig\int d^4x~\bar{\zeta}^{\ad}\pa^{\a}{}_{\ad}\zeta_{\a}\IEEEyesnumber\\
&+m\int d^4x~\phi^{\a}\left\{\zeta_{\a}-\frac{R_{\a}}{m}\right\}+c.c.
\eea
Now we redefine $\psi_{\a}$
\be
\psi_{\a}\to \psi_{\a}+d\phi_{\a}.
\ee
The change of $\mathcal{L}_{h=\pm 3/2}$ is 
\bea{ll}
\delta\mathcal{L}_{h=\pm 3/2}=&-\frac{d}{2\sqrt{2}}\phi^{\a}R_{\a}+c.c.\IEEEyesnumber\\
&-\frac{3}{4}|d|^2i\bar{\phi}^{\ad}\pa^{\a}{}_{\ad}\phi_{\a},
\eea
and the change of $R_{\a}$ is 
\be
\delta R_{\a}=\frac{3\bar{d}}{\sqrt{2}}i\pa_{\a}{}^{\ad}\bar{\phi}_{\ad}.
\ee
So we get that
\bea{ll}
\mathcal{S'}|_{{}_{F}}=\int d^4x\mathcal{L}_{h=\pm 3/2}&+ig\int d^4x~\bar{\zeta}^{\ad}\pa^{\a}{}_{\ad}\zeta_{\a}+m\int d^4x~\left\{\phi^{\a}\zeta_{\a}+\bar{\phi}^{\ad}\bar{\zeta}_{\ad}\right\}\IEEEyesnumber\\
&-\left(\frac{d}{2\sqrt{2}}+1\right)\int d^4x~\phi^{\a}R_{\a}+c.c.\\
&-\left(\frac{3}{4}|d|^2+\frac{6\bar{d}}{\sqrt{2}}\right)i\int d^4x~\bar{\phi}^{\ad}\pa^{\a}{}_{\ad}\phi_{\a}.
\eea
Finally we choose $d$ in order to cancel the interaction term with $R_{\a}$
\be
d+2\sqrt{2}=0,
\ee
and  we get
\bea{ll}
\label{gravitinoHD}
\mathcal{S'}|_{{}_{F}}=\int d^4x\mathcal{L}_{h=\pm 3/2}&+ig\int d^4x~\bar{\zeta}^{\ad}\pa^{\a}{}_{\ad}\zeta_{\a} 
 +6i\int d^4x~\bar{\phi}^{\ad}\pa^{\a}{}_{\ad}\phi_{\a}\IEEEyesnumber\\
&+m\int d^4x~\left\{\phi^{\a}\zeta_{\a}+\bar{\phi}^{\ad}\bar{\zeta}_{\ad}\right\}.
\eea
The equations of motion from Lagrangian 
(\ref{gravitinoHD}) 
on top of the massless gravitino,  
give two massive Weyl spinors with Dirac mass $\mu^2=m^2/(6g)$.
\end{enumerate}
Therefore the spectrum of fermions gives, as expected, the same structure.

%%%%%%%%%%%%%%%%%%%%%%%%%%%%%%%%%%%%%%%%%%%%%%

\subsection{Superspace Duality}

From our previous considerations, we find that this 
higher curvature theory has additional propagating degrees of freedom. 
Since this is a supersymmeric theory it should be possible to identify the multiplet structure of these 
new degrees of freedom directly from superspace manipulations. 
In other words we expect to find that our higher curvature theory 
is classically equivalent to a particular set of matter fields coupled to {\it standard} non-minimal supergravity 
(i.e. a supergravity with no higher curvature terms). 
The Superspace action for the above choice of coefficients is of the form
\bea{lll}
\label{SRR}
\mathcal{S}&=\mathcal{S}_{R}&~-\frac{1}{4}\frac{g}{m^2}\int d^8z G^{\a}\D_{\a}\Dd^{\ad}\bar{G}_{\ad}\\
& &~+\frac{g}{m^2}\int d^8z G^{\a}\Dd^{\ad}\D^{\a}\bar{G}_{\ad}\IEEEyesnumber\\
& &~+\frac{g}{m^2}\int d^8z G^{\a}\Dd^2\bar{G}_{\ad}+c.c.\\
&=\mathcal{S}_{R}&~+\frac{g}{4m^2}\int d^8z \bar{\Phi}\Phi~-\frac{g}{2m^2}\int d^8z V^{\a\ad}V_{\a\ad},
\eea
for the chiral $\Phi$=$\Dd^{\ad}\bar{G}_{\ad}$  
and the real vector $V_{\a\ad}$=$i(\D_{\a}\bar{G}_{\ad}+\Dd_{\ad}G_{\a})$. 
The action (\ref{SRR}) can be re-written as
\bea{lll}
\mathcal{S}&=\mathcal{S}_{R}&+mk\int d^8z T(S-\frac{\Phi}{m})+mk\int d^8z \bar{T}(\bar{S}-\frac{\bar{\Phi}}{m})+\frac{g}{4}\int d^8z \bar{S}S\IEEEyesnumber\\
& &+l\int d^8z F^{\a\ad}V_{\a\ad}+m^2\frac{l^2}{2g}\int d^8z F^{\a\ad}F_{\a\ad},
\eea
where $T$ ([$T$]=$0$) is an unconstrained scalar superfield, $S$ ([$S$]=$1$) is a chiral superfield and $F_{\a\ad}$ ([$F_{\a\ad}$]=$0$) is a real vector superfield.
Indeed, the equations of motion of $T$ and $F_{\a\ad}$ lead to the original action  (\ref{SRR}). 
Now we perform the following shift 
\be
\chi_{\a}\to\chi_{\a}+c\D_{\a}T+id\Dd^{\ad}F_{\a\ad},
\ee
under which we find
%\begin{itemize}
%
%\item For the standard gravitational sector 
%\bea{lll}
%\mathcal{S}_{R}\to &\mathcal{S}_{R}&-c\int d^8z \left\{T\Phi+\bar{T}\bar{\Phi}\right\}+d\int d^8z F^{\a\ad}V_{\a\ad}\\
%& &+4cd \left(\int d^8z T\Dd^2\pa^{\a\ad}F_{\a\ad}+c.c. \right) 
%-8c^2\int d^8z T\Dd^2\D^2\bar{T}\IEEEyesnumber\\
%& &+\frac{d^2}{2}\int d^8z\left\{F^{\a\ad}[\D_{\a},\Dd_{\ad}][\D^{\b},\Dd^{\b}]F_{\b\bd}+3\pa_{\a\ad}\pa^{\b\bd}F_{\b\bd}\right\}.
%\eea
%
%\item For the $T-\Phi$ sector
%\bea{ll}
%\int d^8z T\Phi +c.c.\to \int d^8z T\Phi ~& -4d\int d^8z T\Dd^2\pa^{\a\ad}F_{\a\ad}+c.c.\IEEEyesnumber\\
%&+16kc\int d^8z T\Dd^2\D^2\bar{T}.
%\eea
%
%\item For the $F-V$ sector
%\bea{ll}
%\int d^8z F^{\a\ad}V_{\a\ad}\to \int d^8zF^{\a\ad}V_{\a\ad}+&d\int d^8z\left\{F^{\a\ad}[\D_{\a},\Dd_{\ad}][\D^{\b},\Dd^{\b}]F_{\b\bd}+3\pa_{\a\ad}\pa^{\b\bd}F_{\b\bd}\right\}\\
%&+4c\int d^8z T\Dd^2\pa^{\a\ad}F_{\a\ad}+c.c. \IEEEyesnumber
%\eea
%
%\end{itemize}
%Putting everything together we get 
\bea{lll}
\mathcal{S}=&\mathcal{S}_{R}&+mk\int d^8z\left\{TS+\bar{T}\bar{S}\right\}+\frac{g}{m^2}\int d^8z\bar{S}S+m^2\frac{l^2}{2g}\int d^8z F^{\a\ad}F_{\a\ad}\\
& &-[k+c]\int d^8z \left\{T\Phi+\bar{T}\bar{\Phi}\right\} +[l+d]\int d^8z F^{\a\ad}V_{\a\ad}\IEEEyesnumber\\
& &+[4kd+4cd+4lc]\int d^8z T\Dd^2\pa^{\a\ad}F_{\a\ad}+c.c.\\
& &-[16kc+8c^2]\int d^8z T\Dd^2\D^2\bar{T}\\
& &+[\frac{d^2}{2}+ld]\int d^8z\left\{F^{\a\ad}[\D_{\a},\Dd_{\ad}][\D^{\b},\Dd^{\b}]F_{\b\bd}+3\pa_{\a\ad}\pa^{\b\bd}F_{\b\bd}\right\}. 
\eea 
We now choose coefficients $c$ and $d$ to eliminate the cross terms involving superfields $\Phi$ and $V_{\a\ad}$ respectively, 
which gives $c=-k$ and $d=-l$, 
leading to  
\bea{lll}
\label{standardHD}
\mathcal{S}=&\mathcal{S}_{R}&+mk\int d^8z\left\{TS+\bar{T}\bar{S}\right\}+\frac{g}{m^2}\int d^8z\bar{S}S+m^2\frac{l^2}{2g}\int d^8z F^{\a\ad}F_{\a\ad}\\
& &-4lk \left( \int d^8z T\Dd^2\pa^{\a\ad}F_{\a\ad}+c.c. \right) +8k^2\int d^8z T\Dd^2\D^2\bar{T}\IEEEyesnumber\\
& &-\frac{l^2}{2}\int d^8z\left\{F^{\a\ad}[\D_{\a},\Dd_{\ad}][\D^{\b},\Dd^{\b}]F_{\b\bd}+3\pa_{\a\ad}\pa^{\b\bd}F_{\b\bd}\right\}.
\eea
It is obvious that, the above action contains 
linearized non-minimal supergravity with no higher curvature terms and an independent additional matter sector. Before we conclude let us study the on-shell superfield content of the matter sector, 
and compare to our findings from the component discussion. 

The equations of motion for superfields $F_{\a\ad}$~,~$T$~,~$S$ are
\bea{ll}
\IEEEyesnumber
\label{eq}
\mathcal{E}^{(F)}_{\a\ad}=&-l^2[\D_{\a},\Dd_{\ad}][\D^{\b},\Dd^{\b}]F_{\b\bd}\IEEEyessubnumber -3l^2\pa_{\a\ad}\pa^{\b\bd}F_{\b\bd}\\
&+4lk\pa_{\a\ad}\left(\Dd^2T+\D^2\bar{T}\right) 
+\frac{l^2}{g}m^2F_{\a\ad},\\
{}\\
\mathcal{E}^{T}=&~8k^2\Dd^2\D^2\bar{T} -4lk\Dd^2\pa^{\a\ad}F_{\a\ad} +mkS, \IEEEyessubnumber  \\
{}\\
\mathcal{E}^{S}=&-\frac{g}{4}\Dd^2\bar{S}-mk\Dd^2T.\IEEEyessubnumber
\eea
Looking for the solution of the above equations, we do the following ansatz:
\be
F_{\a\ad}=\pa_{\a\ad}V+[\D_{\a},\Dd_{\ad}]W+\frac{1}{m^2}\pa_{\a\ad}\left(\Dd^2T+\D^2\bar{T}\right), 
\ee
where $V$ and $W$ are on-shell, real,  superfields 
which they satisfy the constraints $D^2V=\Dd^2V=0$, $\D^2W=\Dd^2W=0$ and 
we have for their equations of motion 
\be
\label{eq2}
\D^\g\Dd^2\D_{\g}V+\kappa_{V}mV=0 \ , \  \D^\g\Dd^2\D_{\g}W+\kappa_{W}mW=0.
\ee 
%Note that instead of $V$ we could have used the combination $\D^\a\zeta_{\a}+c.c.$ with $\zeta_{\a}$ to %be a chiral. 
%Then all the above constraints follow. Similar for $W$. 
By doing that, we realize that there are two on-shell chiral supermultiplets, described by the chiral superfields $\Dd^2T$ and $S$ and they satisfy the following equations of motion 
\be
\label{eq3}
\Box(\Dd^2T)=\kappa_{T}m^2(\Dd^2T) \ , \  \Box S=\kappa_{S}m^2 S~.
\ee 
%Therefore the above equations of motion become:
%\bea{ll}
%\IEEEyesnumber
%&-6l^2\kappa_{V}+\frac{l^2}{g}=0\IEEEyessubnumber\\
%&-6l^2\kappa_{W}+\frac{l^2}{g}=0\IEEEyessubnumber\\
%&-4l^2\kappa_{T}+4lk+\frac{l^2}{g}=0\IEEEyessubnumber\\
%&(8k^2-8lk\kappa_{T})\kappa_{T}-\frac{4k^2}{g}=0\IEEEyessubnumber\\
%&-\frac{g}{4}\kappa_{S}+\frac{k^2}{8k^2-8lk\kappa_{T}}=0\IEEEyessubnumber
%\eea
%The solution of the above is 
The above equations (\ref{eq2}) and (\ref{eq3}) solve the system (\ref{eq}) if we set 
\be
\kappa_{V}=\kappa_{W}=\kappa_{T}=\kappa_{S}=\frac{1}{6g}~,~k=-\frac{l}{12g}. 
\ee
From (\ref{eq2}) and (\ref{eq3}) we see that indeed 
we get two vector supermultiplets and two chiral supermultiplets with equal masses $\mu^2=\frac{m^2}{6g}$.
The final expression for the superspace action is
\bea{lll}
\mathcal{S}=&\mathcal{S}_{R}&-\frac{l^2}{12}\frac{m}{g}\int d^8z\left\{TS+\bar{T}\bar{S}\right\}+\frac{g}{m^2}\int d^8z\bar{S}S+m^2\frac{l^2}{2g}\int d^8z F^{\a\ad}F_{\a\ad}\\
& & +\frac{l^2}{18g^2}\int d^8z T\Dd^2\D^2\bar{T}  
+ \left( \frac{l^2}{3g}\int d^8z T\Dd^2\pa^{\a\ad}F_{\a\ad}+c.c. \right) \IEEEyesnumber\\
& &-\frac{l^2}{2}\int d^8z\left\{F^{\a\ad}[\D_{\a},\Dd_{\ad}][\D^{\b},\Dd^{\b}]F_{\b\bd}+3\pa_{\a\ad}\pa^{\b\bd}F_{\b\bd}\right\} 
\eea
where $g$ and $l$ are free, non-zero parameters. 
Furthermore, due to the different integration by parts properties of the two operators $\pa_{\a\ad}$ and $[\D_{\a},\Dd_{\ad}]$, we immediately conclude that there will be an overall minus in front of the terms quadratic to $W$, illustrating that, the $W$ massive vector supermultiplet will be a tachyonic ghost one.  
%This can be simply demonstrated by a straightforward expansion of the %erms in the superspace action quadratic to $F_{\a\ad}$
%\bea{ll}
%&m^2\frac{l^2}{2g}\int d^8z F^{\a\ad}F_{\a\ad}=\int d^8z\left\{-%%m^2\frac{l^2}{g}V\Box V+3m^2\frac{l^2}{g}W\Box W-2\frac{l^2}{m^2g}%\Dd^2T\Box\D^2\bar{T}\right\},\IEEEyesnumber\\
%&-\frac{l^2}{2}\int d^8z F^{\a\ad}[\D_{\a},\Dd_{\ad}][\D^{\b},%\Dd^{\b}]F_{\b\bd}=\int d^8z \left\{-18l^2W\Box\Box W -4\frac{l^2}{m^4}%\Dd^2T\Box\Box\D^2\bar{T}\right\},\IEEEyesnumber\\
%&-\frac{3l^2}{2}\int d^8z F^{\a\ad}\pa_{\a\ad}\pa^{\b\bd}F_{\b\bd}=\int %d^8z \left\{+6l^2V\Box\Box V +12\frac{l^2}{m^4}\Dd^2T\Box\Box%\D^2\bar{T}\right\}\IEEEyesnumber.
%\eea
The above performed superspace  duality demonstrated the classical equivalence between the higher curvature non-minimal supergravity theory and the non-minimal supergravity with the addition\hyphenation{a-ddi-tion} of a specific spectrum that we are expecting from the previous component discussions.

\section{Conclusions}

In this work we have studied the spectrum of the Starobinsky model $R+R^2$, embedded in the framework of non-minimal supergravity. We have utilized the linearized theory since it is sufficient for the understanding of the field content. 
As expected from a supergravity theory, 
on top of the scalaron degree of freedom, 
there are previously auxiliary fields which now pick up kinematic terms due to to the new action. 
We have identified these fields and the way they organize inside supermultiplets. 
Our findings show that the 20/20 higher curvature\hyphenation{cu-rva-tu-re} supergravity is 
classically equivalent to a 20/20 supergravity coupled to 
two vector supermultiplets (one of which is a tachyonic ghost multiplet) and two chiral supermultiplets with equal masses. 
Therefore, the embedding of the $R+R^2$ theory in non-minimal supergravity is reminiscent of the corresponding embedding of the general quadratic gravity (with $R^2$ and Weyl square terms) in minimal supergravity, as in both cases unphysical states appear in the spectrum.

\section*{Acknowledgments}

The work of F.F.  is supported by the Grant agency of the Czech republic under the grant P201/12/G028. 
The  research of A.K. was implemented under the ``Aristeia II" Action of the 
``Operational Programme Education and Lifelong Learning''
and is co-funded by the European 
Social Fund (ESF) and National Resources.  It is partially
supported by European Union's Seventh Framework Programme (FP7/2007-2013) under REA
grant agreement n. 329083.

%%%%%%%%%%%%%%%%%%%%%%%%%%%%%%%%%%%%%%%%%%%%%%

%%%%%%%%%%%%%%%%%%%%%%%%%%%%%%%%%%%%%%%%%%%%%%
%%%%%%%%%%%%%%%%%%%%%%%%%%%%%%%%%%%%%%%%%%%%%%
%%%%  Bibliography %%%%%%%%%%%%%%%%%%%%%%%%%%%

%\bibliography{references}        %%%% File containing the bibtex references %%%%

\begin{thebibliography}{99}


\bibitem{SG} 
  S.~J.~Gates, M.~T.~Grisaru, M.~Rocek and W.~Siegel,
  ``Superspace Or One Thousand and One Lessons in Supersymmetry,''
  hep-th/0108200;\newline
%  %%CITATION = HEP-TH/0108200;%%
  I.~L.~Buchbinder and S.~M.~Kuzenko,
  ``Ideas and methods of supersymmetry and supergravity: A Walk through superspace,''
  Bristol, UK: IOP (1995);\newline
  %4 citations counted in INSPIRE as of 28 Jan 2015
  D.~Z.~Freedman and A.~Van Proeyen,
  ``Supergravity,''
  Cambridge, UK: Cambridge Univ. Pr. (2012)


%\cite{Ade:2013uln}
\bibitem{Ade:2013uln} 
  P.~A.~R.~Ade {\it et al.}  [Planck Collaboration],
  %``Planck 2013 results. XXII. Constraints on inflation,''
  Astron.\ Astrophys.\  {\bf 571}, A22 (2014)\newline
  [arXiv:1303.5082] [astro-ph.CO].
  %%CITATION = ARXIV:1303.5082;%%


%\cite{Lyth:1998xn}
\bibitem{Lyth:1998xn} 
  D.~H.~Lyth and A.~Riotto,
  %``Particle physics models of inflation and the cosmological density perturbation,''
  Phys.\ Rept.\  {\bf 314}, 1 (1999)
  [hep-ph/9807278].
  %%CITATION = HEP-PH/9807278;%%





%\cite{Starobinsky:1980te}
\bibitem{Starobinsky:1980te} 
  A.~A.~Starobinsky,
  %``A New Type of Isotropic Cosmological Models Without Singularity,''
  Phys.\ Lett.\ B {\bf 91}, 99 (1980).
  %%CITATION = PHLTA,B91,99;%%

\bibitem{Stelle} 
  K.~S.~Stelle,
  %``Classical Gravity with Higher Derivatives,''
  Gen.\ Rel.\ Grav.\  {\bf 9}, 353 (1978); 
  %%CITATION = GRGVA,9,353;%%
  %474 citations counted in INSPIRE as of 28 gen 2015
Phys.\ Rev.\ D {\bf 16} (1977) 953.
  %%CITATION = PHRVA,D16,953;%%
  %1044 citations counted in INSPIRE as of 28 Jan 2015

%\cite{Whitt:1984pd}
\bibitem{Whitt:1984pd} 
  B.~Whitt,
  %``Fourth Order Gravity as General Relativity Plus Matter,''
  Phys.\ Lett.\ B {\bf 145}, 176 (1984).
  %%CITATION = PHLTA,B145,176;%%









\bibitem{Wess:1978bu}
  J.~Wess and B.~Zumino,
  %``Superfield Lagrangian for Supergravity,''
  Phys.\ Lett.\ B {\bf 74} (1978) 51;\newline
  K.~S.~Stelle and P.~C.~West,
  %``Minimal Auxiliary Fields for Supergravity,''
  Phys.\ Lett.\ B {\bf 74} (1978) 330;\newline
  S.~Ferrara and P.~van Nieuwenhuizen,
  %``The Auxiliary Fields of Supergravity,''
  Phys.\ Lett.\ B {\bf 74} (1978) 333.

%\cite{Sohnius:1981tp}
\bibitem{west}
  M.~F.~Sohnius and P.~C.~West,
  %``An Alternative Minimal Off-Shell Version of N=1 Supergravity,''
  Phys.\ Lett.\ B {\bf 105}, 353 (1981).
  %%CITATION = PHLTA,B105,353;%%
\bibitem{Howe:1981et}
 P.~S.~Howe, K.~S.~Stelle and P.~K.~Townsend
 Phys.\ Lett. {\bf 107B} 420 (1981);\newline
  S.~J.~Gates, Jr., M.~Rocek and W.~Siegel,
  %``Solution to Constraints for $n=0$ Supergravity,''
  Nucl.\ Phys.\ B {\bf 198} (1982) 113.

\bibitem{Siegel:1978mj}
  W.~Siegel and S.~J.~Gates, Jr.,
  %``Superfield Supergravity,''
  Nucl.\ Phys.\ B {\bf 147} (1979) 77;\newline
  P.~Breitenlohner,
  %``On the Auxiliary Fields of Supergravity,''
  Phys.\ Lett.\ B {\bf 80} (1979) 217.


%\cite{Ferrara:1978rk}
\bibitem{Ferrara:1978rk} 
  S.~Ferrara, M.~T.~Grisaru and P.~van Nieuwenhuizen,
  %``Poincare and Conformal Supergravity Models With Closed Algebras,''
  Nucl.\ Phys.\ B {\bf 138}, 430 (1978).
  %%CITATION = NUPHA,B138,430;%%


%\cite{Cecotti:1987sa}
\bibitem{Cecotti:1987sa} 
  S.~Cecotti,
  %``Higher Derivative Supergravity Is Equivalent To Standard Supergravity Coupled To Matter. 1.,''
  Phys.\ Lett.\ B {\bf 190}, 86 (1987).
  %%CITATION = PHLTA,B190,86;%%


%\cite{Kallosh:2013lkr}
\bibitem{Kallosh:2013lkr} 
  R.~Kallosh and A.~Linde,
  %``Superconformal generalizations of the Starobinsky model,''
  JCAP {\bf 1306}, 028 (2013)
  [arXiv:1306.3214 [hep-th]].
  %%CITATION = ARXIV:1306.3214;%%


%\cite{Farakos:2013cqa}
\bibitem{Farakos:2013cqa} 
  F.~Farakos, A.~Kehagias and A.~Riotto,\newline
  %``On the Starobinsky Model of Inflation from Supergravity,''
  Nucl.\ Phys.\ B {\bf 876}, 187 (2013)
  [arXiv:1307.1137].
  %%CITATION = ARXIV:1307.1137;%%


%\cite{Dalianis:2014aya}
\bibitem{Dalianis:2014aya} 
  I.~Dalianis, F.~Farakos, A.~Kehagias, A.~Riotto and R.~von Unge,
  %``Supersymmetry Breaking and Inflation from Higher Curvature Supergravity,''
  arXiv:1409.8299 [hep-th].
  %%CITATION = ARXIV:1409.8299;%%


%\cite{Cecotti:1987qe}
\bibitem{Cecotti:1987qe} 
  S.~Cecotti, S.~Ferrara, M.~Porrati and S.~Sabharwal,
  %``New Minimal Higher Derivative Supergravity Coupled To Matter,''
  Nucl.\ Phys.\ B {\bf 306}, 160 (1988).
  %%CITATION = NUPHA,B306,160;%%


%\cite{Cecotti:1987qr}
\bibitem{Cecotti:1987qr} 
  S.~Cecotti, S.~Ferrara and L.~Girardello,
  %``Massive Vector Multiplets From Superstrings,''
  Nucl.\ Phys.\ B {\bf 294}, 537 (1987).
  %%CITATION = NUPHA,B294,537;%%

%\cite{Ferrara:2013rsa}
\bibitem{Ferrara:2013rsa} 
  S.~Ferrara, R.~Kallosh, A.~Linde and M.~Porrati,
  %``Minimal Supergravity Models of Inflation,''
  Phys.\ Rev.\ D {\bf 88}, no. 8, 085038 (2013)
  [arXiv:1307.7696 [hep-th]].
  %%CITATION = ARXIV:1307.7696;%%

%\cite{Ferrara:2014cca}
\bibitem{Ferrara:2014cca} 
  S.~Ferrara and M.~Porrati,
  %``Minimal $R+R^2$ Supergravity Models of Inflation Coupled to Matter,''
  Phys.\ Lett.\ B {\bf 737}, 135 (2014)
  [arXiv:1407.6164 [hep-th]].
  %%CITATION = ARXIV:1407.6164;%%


%\cite{Ellis:2013xoa}
\bibitem{Ellis:2013xoa} 
  J.~Ellis, D.~V.~Nanopoulos and K.~A.~Olive,
  %``No-Scale Supergravity Realization of the Starobinsky Model of Inflation,''
  Phys.\ Rev.\ Lett.\  {\bf 111}, 111301 (2013)
  [Erratum-ibid.\  {\bf 111}, no. 12, 129902 (2013)]
  [arXiv:1305.1247 [hep-th]].
  %%CITATION = ARXIV:1305.1247;%%

%\cite{Ellis:2013nxa}
\bibitem{Ellis:2013nxa} 
  J.~Ellis, D.~V.~Nanopoulos and K.~A.~Olive,\\
  %``Starobinsky-like Inflationary Models as Avatars of No-Scale Supergravity,''
  JCAP {\bf 1310}, 009 (2013)
  [arXiv:1307.3537].
  %%CITATION = ARXIV:1307.3537;%%

%\cite{Ferrara:2013eqa}
\bibitem{Ferrara:2013eqa} 
  S.~Ferrara, P.~Fre and A.~S.~Sorin,
  %``On the Topology of the Inflaton Field in Minimal Supergravity Models,''
  JHEP {\bf 1404}, 095 (2014)
  [arXiv:1311.5059 [hep-th]].
  %%CITATION = ARXIV:1311.5059;%%

%\cite{Ferrara:2014rya}
\bibitem{Ferrara:2014rya} 
  S.~Ferrara, P.~Fre and A.~S.~Sorin,
  %``On the Gauged KÄÂ¤hler Isometry in Minimal Supergravity Models of Inflation,''
  Fortsch.\ Phys.\  {\bf 62}, 277 (2014)
  [arXiv:1401.1201 [hep-th]].
  %%CITATION = ARXIV:1401.1201;%%

%\cite{Antoniadis:2014oya}
\bibitem{Antoniadis:2014oya} 
  I.~Antoniadis, E.~Dudas, S.~Ferrara and A.~Sagnotti,
  %``The Volkov-Akulov-Starobinsky supergravity,''
  Phys.\ Lett.\ B {\bf 733}, 32 (2014)
  [arXiv:1403.3269 [hep-th]].
  %%CITATION = ARXIV:1403.3269;%%


%\cite{Farakos:2014gba}
\bibitem{Farakos:2014gba} 
  F.~Farakos and R.~von Unge,
  %``Naturalness and Chaotic Inflation in Supergravity from Massive Vector Multiplets,''
  JHEP {\bf 1408}, 168 (2014)
  [arXiv:1404.3739 [hep-th]].
  %%CITATION = ARXIV:1404.3739;%%


%\cite{Ellis:2014gxa}
\bibitem{Ellis:2014gxa} 
  J.~Ellis, M.~A.~G.~Garcia, D.~V.~Nanopoulos and K.~A.~Olive,
  %``A No-Scale Inflationary Model to Fit Them All,''
  JCAP {\bf 1408}, 044 (2014)
  [arXiv:1405.0271 [hep-ph]].
  %%CITATION = ARXIV:1405.0271;%%


%\cite{Ketov:2014qha}
\bibitem{Ketov:2014qha} 
  S.~V.~Ketov and T.~Terada,
  %``Inflation in Supergravity with a Single Chiral Superfield,''
  Phys.\ Lett.\ B {\bf 736}, 272 (2014)
  [arXiv:1406.0252 [hep-th]].
  %%CITATION = ARXIV:1406.0252;%%


%\cite{Ferrara:2014yna}
\bibitem{Ferrara:2014yna} 
  S.~Ferrara and A.~Kehagias,
  %``Higher Curvature Supergravity, Supersymmetry Breaking and Inflation,''
  arXiv:1407.5187 [hep-th].
  %%CITATION = ARXIV:1407.5187;%%


%\cite{Ferrara:2014kva}
\bibitem{Ferrara:2014kva} 
  S.~Ferrara, R.~Kallosh and A.~Linde,
  %``Cosmology with Nilpotent Superfields,''
  JHEP {\bf 1410}, 143 (2014)
  [arXiv:1408.4096 [hep-th]].
  %%CITATION = ARXIV:1408.4096;%%


%\cite{Ketov:2014hya}
\bibitem{Ketov:2014hya} 
  S.~V.~Ketov and T.~Terada,
  %``Generic Scalar Potentials for Inflation in Supergravity with a Single Chiral Superfield,''
  JHEP {\bf 1412}, 062 (2014)
  [arXiv:1408.6524 [hep-th]].
  %%CITATION = ARXIV:1408.6524;%%


%\cite{Dall'Agata:2014oka}
\bibitem{Dall'Agata:2014oka} 
  G.~Dall'Agata and F.~Zwirner,
  %``On sgoldstino-less supergravity models of inflation,''
  JHEP {\bf 1412}, 172 (2014)
  [arXiv:1411.2605 [hep-th]].
  %%CITATION = ARXIV:1411.2605;%%


%\cite{Terada:2014uia}
\bibitem{Terada:2014uia} 
  T.~Terada, Y.~Watanabe, Y.~Yamada and J.~Yokoyama,
  %``Reheating processes after Starobinsky inflation in old-minimal supergravity,''
  arXiv:1411.6746 [hep-ph].
  %%CITATION = ARXIV:1411.6746;%%


%\cite{Alexandre:2013nqa}
\bibitem{Alexandre:2013nqa} 
  J.~Alexandre, N.~Houston and N.~E.~Mavromatos,
  %``Starobinsky-type Inflation in Dynamical Supergravity Breaking Scenarios,''
  Phys.\ Rev.\ D {\bf 89}, no. 2, 027703 (2014)
  [arXiv:1312.5197 [gr-qc]].
  %%CITATION = ARXIV:1312.5197;%%



%\cite{Alexandre:2014lla}
\bibitem{Alexandre:2014lla} 
  J.~Alexandre, N.~Houston and N.~E.~Mavromatos,
  %``Inflation via Gravitino Condensation in Dynamically Broken Supergravity,''
  arXiv:1409.3183 [gr-qc].
  %%CITATION = ARXIV:1409.3183;%%


%\cite{Lang:1985xk}
\bibitem{Lang:1985xk} 
  W.~Lang, J.~Louis and B.~A.~Ovrut,
  %``(16+16) Supergravity Coupled to Matter: The Low-energy Limit of the Superstring,''
  Phys.\ Lett.\ B {\bf 158}, 40 (1985).
  %%CITATION = PHLTA,B158,40;%%


%\cite{Hayashi:1985vd}
\bibitem{Hayashi:1985vd} 
  M.~Hayashi and S.~Uehara,
  %``On The 'new' (16+16) Version Of N=1 Supergravity,''
  Phys.\ Lett.\ B {\bf 172}, 348 (1986).
  %%CITATION = PHLTA,B172,348;%%


%\cite{Aulakh:1985dn}
\bibitem{Aulakh:1985dn}
  C.~S.~Aulakh, J.~P.~Derendinger and S.~Ouvry,
  %``On the Reducibility of (16+16) Supergravity,''
  Phys.\ Lett.\ B {\bf 169} (1986) 201.
  %%CITATION = PHLTA,B169,201;%%

%\cite{Siegel:1986sv}
\bibitem{Siegel:1986sv} 
  W.~Siegel,
  %``16/16 Supergravity,''
  Class.\ Quant.\ Grav.\  {\bf 3}, L47 (1986).
  %%CITATION = CQGRD,3,L47;%%


%%\cite{Breitenlohner:1976nv}
%\bibitem{Breitenlohner:1976nv} 
%  P.~Breitenlohner,
%  %``A Geometric Interpretation of Local Supersymmetry,''
%  Phys.\ Lett.\ B {\bf 67}, 49 (1977).
%  %%CITATION = PHLTA,B67,49;%%
%
%
%
%%\cite{Breitenlohner:1977jn}
%\bibitem{Breitenlohner:1977jn} 
%  P.~Breitenlohner,
%  %``Some Invariant Lagrangians for Local Supersymmetry,''
%  Nucl.\ Phys.\ B {\bf 124}, 500 (1977).
%  %%CITATION = NUPHA,B124,500;%%
%
%
%%\cite{Girardi:1984eq}
%\bibitem{Girardi:1984eq} 
%  G.~Girardi, R.~Grimm, M.~Muller and J.~Wess,
%  %``Superspace Geometry and the Minimal, Nonminimal, and New Minimal Supergravity Multiplets,''
%  Z.\ Phys.\ C {\bf 26}, 123 (1984).
%  %%CITATION = ZEPYA,C26,123;%%
%
%
%%\cite{Grimm:1984pj}
%\bibitem{Grimm:1984pj} 
%  R.~Grimm, M.~Muller and J.~Wess,
%  %``Supersymmetric Gauge Theories Coupled to Nonminimal and New Minimal Supergravity,''
%  Z.\ Phys.\ C {\bf 26}, 427 (1984).
%  %%CITATION = ZEPYA,C26,427;%%
%
%
%%\cite{Garreis:1987cv}
%\bibitem{Garreis:1987cv} 
%  R.~Garreis and C.~Schwiebert,
%  %``Superhiggs Effect In N=1 Nonminimal Supergravity,''
%  Nucl.\ Phys.\ B {\bf 296}, 902 (1988).
%  %%CITATION = NUPHA,B296,902;%%

\bibitem{Gates:2014tea}
  S.~J.~Gates and K.~Koutrolikos,
  %``On 4D, $\mathcal{N} = 1$ massless gauge superďŹelds of arbitrary superhelicity,''
  JHEP {\bf 1406} (2014) 098;\newline
  Gates, S. James., Jr. and K.~Koutrolikos,
  %``On 4D, N = 1 Massless Gauge Superfields of Higher Superspin: Half-Odd-Integer Case,''
  arXiv:1310.7386 [hep-th];\newline
  S.~J.~Gates, Jr. and K.~Koutrolikos,
  %``On 4D, N = 1 Massless Gauge Superfields of Higher Superspin: Integer Case,''
  arXiv:1310.7385 [hep-th].
  
\bibitem{Kuzenko:1993jq}
  S.~M.~Kuzenko and A.~G.~Sibiryakov,
  %``Massless gauge superfields of higher integer superspins,''
  JETP Lett.\  {\bf 57} (1993) 539
   [Pisma Zh.\ Eksp.\ Teor.\ Fiz.\  {\bf 57} (1993) 526].
   
\bibitem{Kuzenko:1993jp}
  S.~M.~Kuzenko, A.~G.~Sibiryakov and V.~V.~Postnikov,
  %``Massless gauge superfields of higher half integer superspins,''
  JETP Lett.\  {\bf 57} (1993) 534
   [Pisma Zh.\ Eksp.\ Teor.\ Fiz.\  {\bf 57} (1993) 521].

%\bibitem{Gates:2011qa}
%  S.~J.~Gates, Jr. and K.~Koutrolikos,
%  %``A Codicil To Massless Gauge Superfields of Higher Half-Odd Integer Superspins,''
%  arXiv:1103.3564 [hep-th].
  
%\bibitem{Gates:2011qb}
%  S.~J.~Gates, Jr. and K.~Koutrolikos,
%  %``A Codicil To Massless Gauge Superfields of Higher Integer Superspins,''
%  arXiv:1103.3565 [hep-th].


%
%\bibitem{Codi} 
% S.\ J.\ Gates, Jr.\, and K. Koutrolikos, ``A Codicil To Massless Gauge 
% Superfields of Higher Half-Odd Integer Superspins,ÄÂ arXiv:1103.3564 [hep-th];
% ibid.\ ``A Codicil To Massless Gauge Superfields of Higher Integer
%Superspins,'' arXiv:1103.3565 [hep-th].
% 
%
%\bibitem{GrvTn} 
%S.\ J.\ Gates, Jr., and W.\ Siegel, W, ``(3/2, 1) Superfield of O(2) Supergravity,''
%Nucl.\ Phys. {\bf {B164}} (1980) 484.

\end{thebibliography}
%\bibliographystyle{unsrt}

                                 %%%% OR

%\newpage

%%%%%%%%%%%%%%%%%%%%%%%%%%%%%%%%%%%%%%%%%%%%%%
%%%%%%%%%%%%%%%%%%%%%%%%%%%%%%%%%%%%%%%%%%%%%%
%%%%  END --- DOCUMENT %%%%%%%%%%%%%%%%%%%%%%%

\end{document}